# High magnetic field induced charge density wave states in a quasi-one dimensional organic conductor

*12/5/03 16:18:22 PM*

D. Graf[a], E.S. Choi[a], J. S. Brooks[a], Rui T. Henriques[b,c], M. Almeida[b], and M. Matos[d]

[a]NHMFL/Physics, Florida State University, Tallahassee, FL 32310, USA

[b]Dept. de Quimica, Instituto-Tecnologico e Nuclear, P-2686-953 Sacavem, Portugal

[c]Instituto de Telecomunicacoes, Instituto Superior Técnico, P-1049-001 Lisboa, Portugal

[d]Dept. de Engenharia Quimica, Instituo Superiore de Engenharia de Lisboa, P-1900 Lisboa, Portugal

We have measured the high field magnetoresistence and magnetization of quasi-one-dimensional (Q1D) organic conductor $(Per)_2Pt(mnt)_2$ (where Per = perylene and mnt = maleonitriledithiolate), which has a charge density wave (CDW) ground state at zero magnetic field below 8 K. We find that the CDW ground state is suppressed with moderate magnetic fields of order 20 T, as expected from a mean field theory treatment of Pauli effects[W. Dieterich and P. Fulde, Z. Physik **265**, 239 - 243 (1973)]. At higher magnetic fields, a new, density wave state with sub-phases is observed in the range 20 to 50 T, which is reminiscent of the cascade of field induced, quantized, spin density wave phases (FISDW) observed in the Bechgaard salts. The new density wave state, which we tentatively identify as a field induced charge density wave state (FICDW), is re-entrant to a low resistance state at even higher fields, of order 50 T and above. Unlike the FISDW ground state, the FICDW state is only weakly orbital, and appears for all directions of magnetic field. Our findings are substantiated by electrical resistivity, magnetization, thermoelectric, and Hall measurements. We discuss our results in light of theoretical work involving magnetic field dependent Q1D CDW ground states in high magnetic fields [D. Zanchi, A. Bjelis, and G. Montambaux, Phys. Rev. B **53**, (1996)1240; A. Lebed, JETP Lett. 78,138(2003)].

**Introduction**

Magnetic field induced spin density wave states (FISDW) have been well established for many years.[1] Here, based on consideration of an imperfectly nested quasi-one-dimensional Q1D Fermi surface, high magnetic fields will reduce the transverse orbital motion and drive the system into a sufficiently one dimensional configuration to allow a spin density wave instability to be stabilized. The resulting FISDW state has sub-phases which are approximately quantized. For a Q1D sytem, $t_x$ is the intra-chain bandwidth, and $t_y$ is the inter-chain bandwidth. Typically, $t_y \ll t_x$. The transverse bandwidth $t_y$ is critical to the FISDW mechanism, due to the Q1D Fermi surface topology, and therefore the resulting FISDW transitions are dependent on field direction, according to $1/\cos(\theta)$, were $\theta$ is the polar angle between the normal to the most conducting x-y plane.

In contrast, charge density wave (CDW) systems have not been widely studied in accessible magnetic fields due to the generally high CDW transition temperatures (of order 100 K or more) which prohibit a comparable Zeeman energy corresponding to fields in excess of 100 T, based on mean field arguments.[2] However, the Q1D system Per$_2$Pt(mnt)$_2$ has a CDW transition temperature of only 9 K.[3] Unlike the FISDW states, the zero field CDW ground state is an insulating density wave state, and the prospects of a field induced CDW are therefore complex. Nevertheless, in recent theoretical work, the magnetic field dependence of a Q1D CDW ground state has been considered.[4] Here the evolution of the native CDW$_0$ ground state, with increasing magnetic field, into an isotropic CDW$_x$ phase and an oribitally dependent CDW$_y$ phase has been predicted, where both spin and orbital terms were includend in a Q1D Hubbard model. For imperfect nesting, quantized cascades of FICDW transitions are also predicted. Indeed, recent work on the Q2D system α-(BEDT-TTF)$_2$KHg(SCN)$_4$ where a similar CDW ground state is expected below 8 K, shows a remarkable similarity with the theoretical expectations, including a cascade of FICDW phases.[5] Here a Q2D Fermi surface co-exists with the Q1D electronic structure, which therefore may complicate comparison with the assumptions of the Q1D theory.

In the case of Per$_2$Pt(mnt)$_2$ however, comparison with the theory for FICDW should be more direct, since the material is highly one dimensional, with only a weak inter-chain bandwidth, nearly an order of magnitude less that of the Bechgaard salts which provide a standard for the properties of Q1D metals. We have therefore investigated the Per$_2$Pt(mnt)$_2$ material in high magnetic fields systematically as a function of temperature and magnetic field direction. We will show that in this highly Q1D CDW system, there is indeed a transition from a CDW$_0$ ground state to a high field insulating state, with discrete sub-phases, which is moderately dependent on field direction. In the high field limit, the insulating state is reduced to a low resistance state.

To preface the following sections, we note that we have recently completed a similar study on the (Per)$_2$Au(mnt)$_2$ system[6], where we found that the magnetic field also suppresses the ambient CDW ground state above 30 T where the results could be reasonably described by mean field theory [2] below 25 T. Although not as dramatic as the Per$_2$Pt(mnt)$_2$ results presented herein, there was also evidence for a new high field induced state above 40 T in accord with the FICDW theory[4]. For brevity, and also for a more detailed discussion of these materials, we refer the reader to Ref. [6].

**Experimental procedures**

Samples of (Per)$_2$Pt(mnt)$_2$ were grown electrochemically [7]. It is essential to note that the CDW transition temperatures of these materials is sensitive to crystal quality, cooling rate dependence, and perhaps to the strain of the electrical contact leads. Hence many different crystals from different synthesis batches were used in the measurements reported here. Moreover, due to the very small size of the crystals, it is not possible to obtain reliable x-ray data from samples used in some of the transport studies. We therefore have adopted the following convention, based on x-ray studies of larger crystals: the b-axis is the very prominent needle axis direction; the c-axis of the crystals is that normal to the widest plane of the needle-like crystals; the a-axis is taken as that normal to the thinnest sample dimension.

Transport and magnetization measurements in dc fields of 33 and 45 T were made at the NHMFL-Tallahassee Lab and pulsed field experiments (6 ms rise-time) were carried out to 60 T at NHMFL-Los Alamos Lab. The resistance of the needle-like single crystals were measured using a standard four-terminal configuration of 12 μm Au wire with currents of 1μA (dc fields) and 10 1μA (pulsed fields). In all cases, the current was applied along the b-axis of the samples. Angular dependence was investigated with a rotating low temperature sample platform stage. Measurements of magnetization used a PRC-120 piezo-resistive cantilever with a small (< 200 μg) $(Per)_2Pt(mnt)_2$ crystal attached to the sample arm. The b-axis (conducting chain) of the crystal was aligned with the rotation axis of the cantilever, thereby tilting the applied field in the ac-plane of the crystal. The cantilever, consisting of both a sample arm and reference arm of equivalent resistance values, is used as part of a bridge circuit, which is balanced at B = 0T.

**Results**

The general behavior of the magnetoresistance of $(Per)_2Pt(mnt)_2$ at low temperatures below the CDW transition at about 8 K is shown in Fig. 1. We find that the ambient CDW phase is suppressed with increasing magnetic field, and that the resistivity decreases from an insulating value to between about 5 to 0.05 Ωcm for 0.5 K. in the 20 to 30 T range. (At low fields, the resistance is not measurable until the CDW is sufficiently suppressed.) At higher fields, a second high resistivity state is entered, where step-like features in the magnetoresistance appear. Finally, above 40 T the second high resistance state is suppressed, and a second low resistance state is entered. In Fig. 1a results for B//a are shown for fields up to 33 T below 4.2 K. In Fig. 1b we show results to 45 T for temperatures below 4.2 K. We note that for this orientation, the magnetic field is parallel to the conducting perylene chains. Here we have indicated a number of features in the data that represent structure in the magnetoresistance, which is common to most samples and orientations covered. We define $B_0$ as the upper critical field of the ambient CDW phase ($CDW_0$). $B_x$ is defined as the onset of the second high resistance phase that we tentatively identify as the field induced $CDW_x$, or FICDW phase. $B_a$, $B_b$, etc. are features that appear to be common to different sample and different field orientation configurations. $B_e$ is the estimated upper critical field of the FICDW. In Fig. 1c, data for the B//c orientation is given up to 45 T, where temperatures approaching the ambient CDW transition temperature are approached. We note that two additional features (slope changes) are seen in the data, one around 21 T at low temperatures, and two more above 35 T as the FICDW state is starting to be suppressed. In Fig. 1c, we note that the high field resistance is still activated ($E_a$ is of order 30 K at 45 T), but its significantly less than the activation energy at the center of the FICDW ($E_a = 80$ K) near 32 T. The drop in resistance near 23 T for low temperatures is not presently understood, although it was reproducible.

The CDW and FICDW phase diagrams derived from the magnetoresistance data of Fig. 1 are shown in Fig. 2. Here we have used the symbol-legend defined in Fig. 1b to demarcate the features common to the magnetoresistance structure in all three magnetic field directions, where we find that the CDW phase is suppressed at high fields, that a FICDW phase is then stabilized with sub-phase structure, and that the FICDW is eventually also suppressed at even higher fields. In particular, the temperature dependence of the sub-phase structure follows a positive slope below (solid symbols), and a negative slope above (open symbols) the center of the FICDW phase. As noted above, we find some additional sub-phase structure near the upper FICDW critical field for B//c in Fig. 1c (and Fig. 2c). Figs. 1 and 2 show that the FICDW, although dependent on field orientation, does not follow a $1/\cos(\theta)$ type behavior as would be expected for a quasi-two-dimensional closed orbit Fermi surface effect.

A test for thermodynamic signatures in the FICDW process have been carried out by torque magnetization measurements to 33 T at 0.5 K for a rotation of the magnetic field in the a-c plane, as shown in Fig. 3. Here B//a corresponds to $0^o$, and B//c corresponds to $90^o$. Fig. 3a shows selected magnetization signals (torque signal divided by magnetic field) for different sample orientations in the a-c plane, where we note that the sign, and also the magnitude of the response is dependent on the sample and cantilever orientation. This is a characteristic of the cantilever response. By inspection, above a smooth background we observe two peaks 1 and 2, which correspond to transitions in the FICDW (which we can only tentatively identify as corresponding to the $B_0$, $B_x$, or $B_a$ transitions, since resistance was not simultaneously measured in this study). By analyzing the second derivative of the torque signal, we find additional peaks, as shown in Fig. 3b. Of note is the peak 3 which has an opposite phase dependence on angle, and the angular dependence of the three peaks is summarized in Fig. 3c. As in the magnetoresistance study, there is clearly no Q2D $1/\cos(\theta)$ type dependence to the field position of the FICDW transitions.

A test of the field orientation anistotropy where a sample was rotated from the B//b orientation to the B//a-c plane direction is shown in Fig. 4. This further confirms that even for the field parallel to the perylene chains, the FICDW appears above the CDW after an intervening low resistance state. As the B//a-c direction is approached, both $B_0$ and $B_x$ decrease in field, but Bx has a significantly larger angular dependence.

Some final experimental tests have been carried out on the FICDW phenomenon. In Fig. 5 we show a comparison of the dc field data to 45 T from Figs. 1 and 4, and also data from a pulsed field study, the purpose of which was to check for additional field induced phases. We find that, at least to 60 T for B//c, there is no additional phase transition. We have measured the thermopower in the FICDW phase, and find it has a conventional gapped behavior, and that the sub-phases appear as structure in the thermopower signal. The Hall effect is also observable in the FICDW, and exhibits structure in the sub-phases. These last two results will be presented elsewhere.[8]

**Discussion and conclusions**

The primary result of the present work is that the charge density wave ground state in a highly one-dimensional conductor is suppressed at moderate magnetic fields; that a second, re-entrant <u>field induced</u> charge density wave state is stabilized at higher fields; and that ultimately even the field induced state is suppressed at even higher fields. The nature of the FICDW shows, through magnetization measurements and also thermopower studies, that it is a bulk thermodynamic process.

The observation this phenomenon is only moderately dependent on magnetic field direction is consistent with the theoretical description of the field dependence given by Zanchi et al.[4], where above the ambient $CDW_0$ phase, a new $CDW_x$ phase is stabilized, which does not depend on field orientation. However, a second phase, the so-called $CDW_y$ phase does involve field orientation through orbital coupling. (Our observation of a Hall signal in the FICDW state is a further indication that an orbital contribution is present, which may even involve nested Q1D pockets.) Another compelling feature of the theory is that for imperfect nesting, a cascade of transitions appear in the $CDW_x$ description, and that in the high field limit, the $CDW_x$ state is suppressed. Likewise, Lebed predicts a cascade of FICDW transitions in a Q1D metal under certain conditions[9]. Clearly, these features appear in our results, but a direct, one-to-one correspondence between our results and the theoretical model are not obvious at this point beyond our qualitative comparison.

The FICDW sub-phases appear in general for all three directions, although some features are evident in some cases. The increasing and decreasing field traces in Fig. 1c show the level of hysteresis typically seen in the FICDW state. We do not find any direct correspondence between the sub-phases and quantum indices, although it is suggestive in some of the traces. It may be that for very high quality samples for specific directions (Lebed predicts magic angle effects[9]), and perhaps at very low temperatures, quantization might be realized, since, as mentioned above, in preliminary studies we do see a Hall signal associated with the FICDW sub-phase structure.

There are many open questions and much work still left to do to fully understand this new FICDW system. As noted in the results, the high field state above 45 T appears activated, at least below about 2 K. This may infer that there is still a gapped, non-metallic state, which may persist to the highest fields. In this case, we would speculate that perhaps a spin-density wave state remains after the complex $CDW_x$ state is suppressed. Temperature dependent pulsed field studies would be useful to check this point. A second point is the necessary condition for an FICDW, i.e., can it evolve from a Q1D metal, or is a well nested CDW ground state a necessary condition. Our results show that the CDW seems to vanish, or nearly vanish, before the FICDW state is stabilized.

**Acknowledgments**

This work is supported by NSF 02-03532, and the NHMFL is supported by a contractual agreement between theNational Science Foundation and the State of Florida. DG is supported by a NSF GK-12 Fellowship. Work in Portugal is supported by FCT under contract POCT/FAT/39115/2001. We are grateful to E. Canadell, G. Bonfait, M. Kartsovnik, and A. Lebed for helpful discussions. We thank the High Field Magnet staff at both Tallahassee and Los Alamos for making these measurements possible.

FIGURE CAPTIONS

Figure 1. Per$_2$Pt(mnt)$_2$ sample resistance signal (voltage for a current of 1 μA) along the b-axis vs. magnetic field for three different crystallographic directions. In all cases the highly resistive (insulating) CDW state is removed above a field of order 20 T. After an intervening low resistance region, the resistance again increases in steps towards a second highly insulating state, which we interpret as a field induced charge density wave (FICDW) state. For fields above 40T, the high resistance state is suppressed. Due to phase shifts that result from the diverging ac resistance signal at low fields, only results where the out of phase component of the signal was less than 10% are shown. a) B//a for fields up to 33T. Here the highest field low resistance state could not be accessed. b) B//b up to 45 T. In this panel we have defined the various steps that appear in the field dependence resistance, in particular, on a logarithmic scale. $B_0$ marks the suppression of the low field CDW state, $B_x$ marks the onset of the new high field insulating state, and $B_a$, $B_b$, etc., mark various features that are common to all samples and field directions. c) B//c up to 45 T. Here we observe three features not evident in the other two cases. One change of slope occurs at a field below what we term $B_0$ in the CDW phase, and the other two occur above $B_d$ in the region 35 to 40 T. The onset of low resistance around 24 T for decreasing temperature is not presently understood, but it is possibly an artifact of the ac measurement. Of note is that at highest field, 45 T, the resistance is still activated, although the resistance of the intervening state has dropped by 4 to 5 orders of magnitude. Of note in ( c ) is the inclusion of field sweeps in both directions for constant temperature, where we find a slight hysteresis in the FICDW phase region.

Figure 2. Proposed magnetic field dependent phase diagram of Per$_2$Pt(mnt)$_2$ based on the magnetotransport studies in Fig. 1. For all three field directions, structure in the magnetoresistance that appear to be common in the FICDW phase, have been plotted as phase and sub-phase boundaries. Closed symbols have been used for transitions below the center of the FICDW, and open symbols are used above the FICDW center. a) B//a phase diagram based on Fig. 1 a).; b) B//b phase diagram based on Fig. 1b). Note that the dashed line for the upper phase boundary $B_e$ is an estimate based on an extrapolation of the data to fields above 45 T. ; c) B//c phase diagram based on Fig. 1c).

Figure 3. Magnetization of Per$_2$Pt(mnt)$_2$ at 0.5 K up to 33 T. a) Torque magnetization signal for a sample rotated in the a-c plane with respect the magnetic field. Although the sign of the response is dependent on the orientation of the cantilever, which rotates in field with the sample, we find at least two clear transitions (**1** and **2**) in the field range 20 to 25 T which we believe are associated with the FICDW phase transitions. b) The second derivative of the torque signals vs. angle. The sequential curves have been systematically offset to reveal the topology of the angular dependent signal. We find that there is additional structure in the signal, including a third peak **3** that appears in a certain range of orientation. c) a-c plane angular dependence of the main magnetization signals. Peaks **1** and **2** follow a simple sinusoidal dependence. Although peak **3** is also sinusoidal, its phase is opposite. These results show that the FICDW phases do not follow a $1/\cos(\phi)$ dependence, as they do in a FISDW system, and therefore the FICDW is not primarily related to a Q2D Fermi surface mechanism.

Figure 4. Angular dependence of the FICDW phase in Per$_2$Pt(mnt)$_2$ for B//b ($\theta = 0^o$) progressing to B//a-c ($\theta \rightarrow 90^o$) at 0.5 K. Inset, polar angle dependence of the B$_{CDW0}$ and B$_{CDWx}$ transitions, as defined in the main panel. As in Fig. 3, we find that the FICDW state does not follow a 1/cos($\theta$) relation, and is not directly related to a Q2D Fermi surface effect.

Figure 5. Pulsed magnetic field study of the FICDW to 60 T at 0.5 K (solid thick line). The FICDW is clearly re-entrant to a high field, low resistance state. Data from Fig. 1 for three different field directions for dc fields are also shown (note linear scale) for comparison. The results indicate that for all orientations, the FICDW terminates before 60 T, although the resistance of the high field re-entrant state may still be weakly activated.

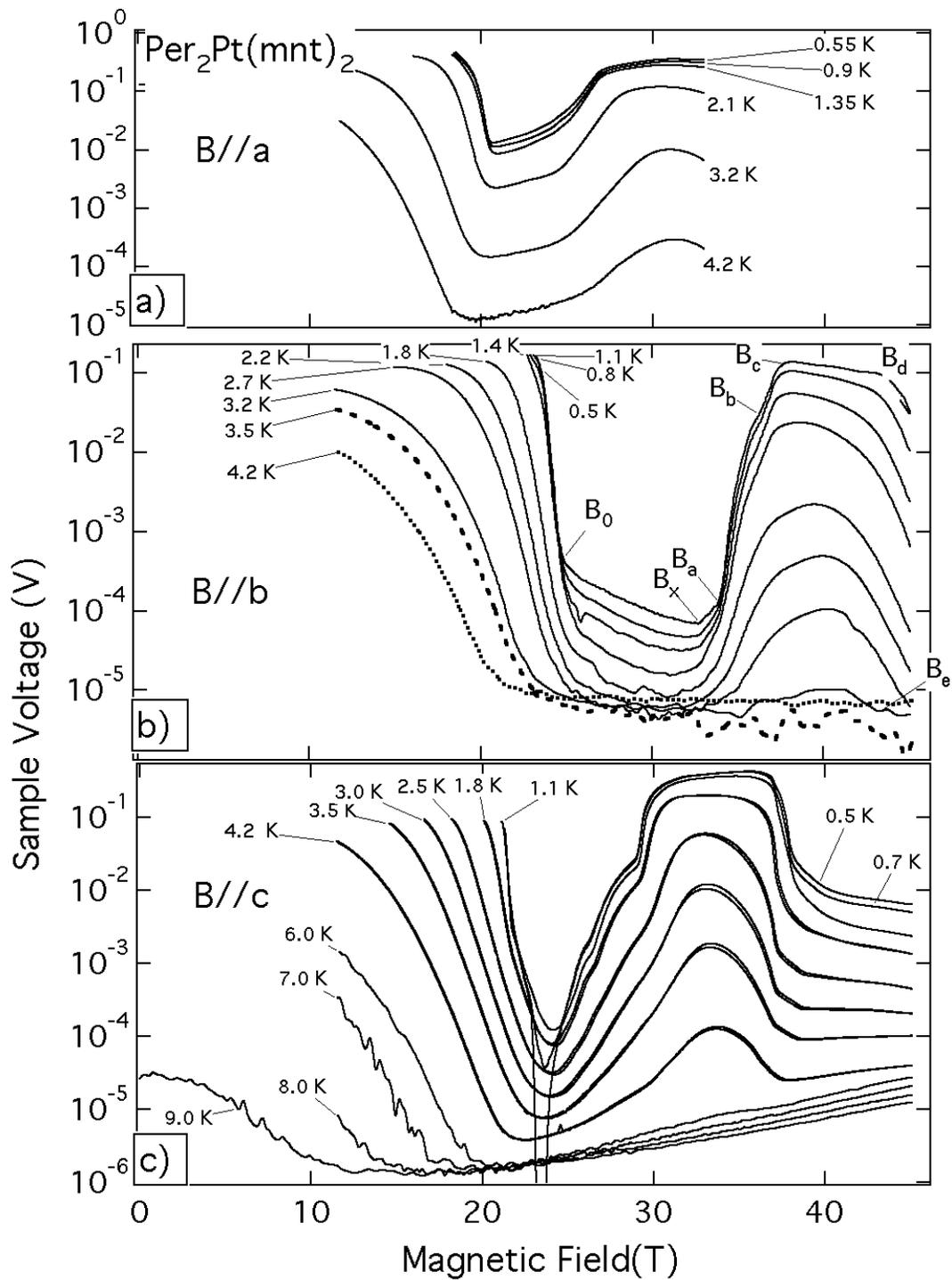

**Figure 1 Graf et al.**

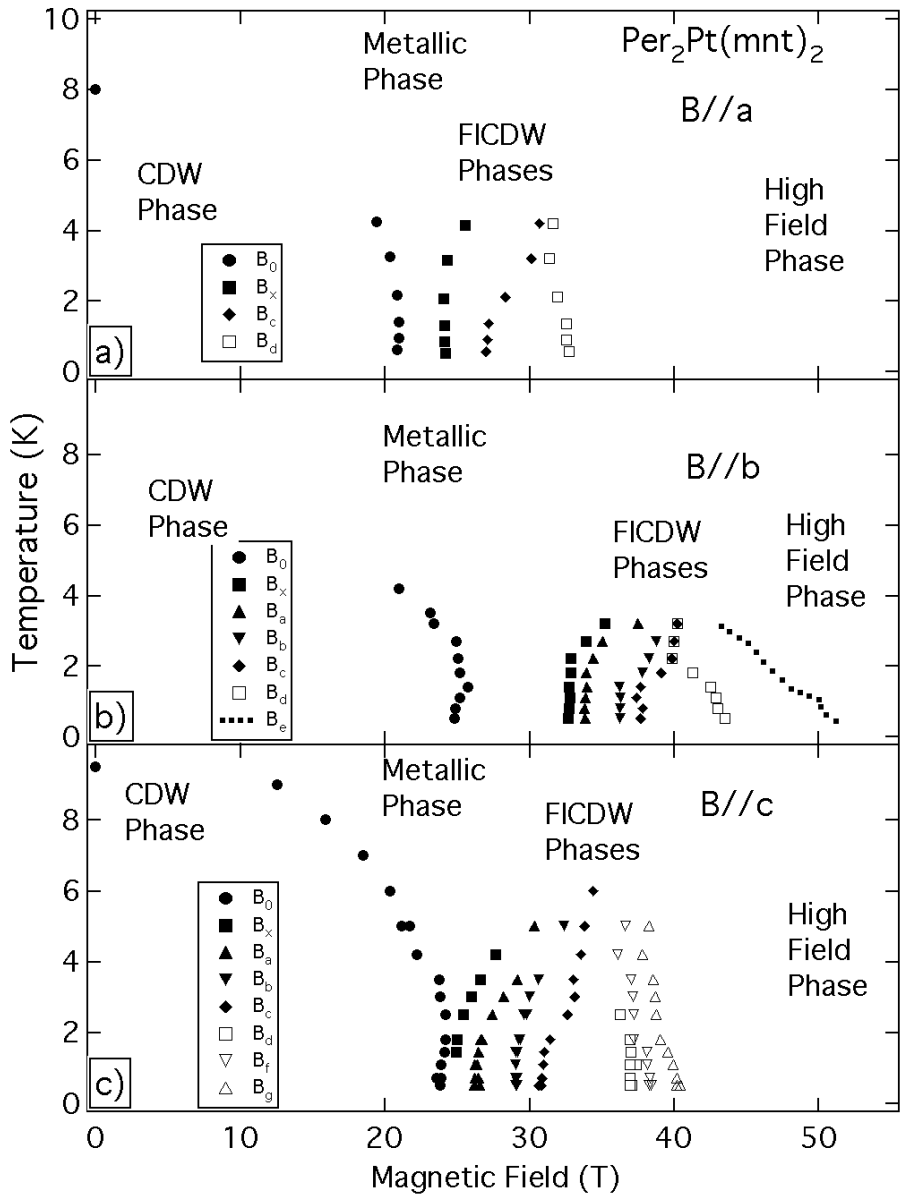

**Figure 2 Graf et al.**

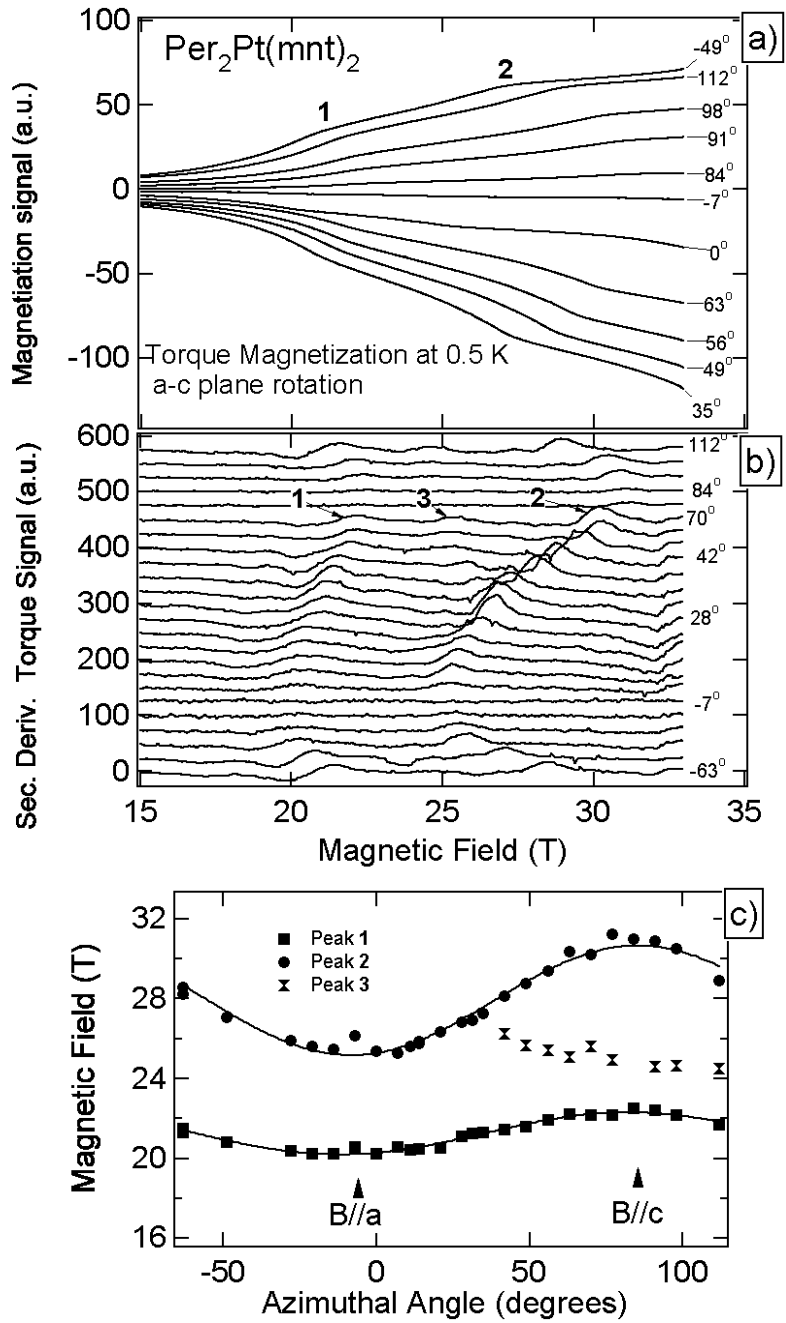

**Figure 3 Graf et al.**

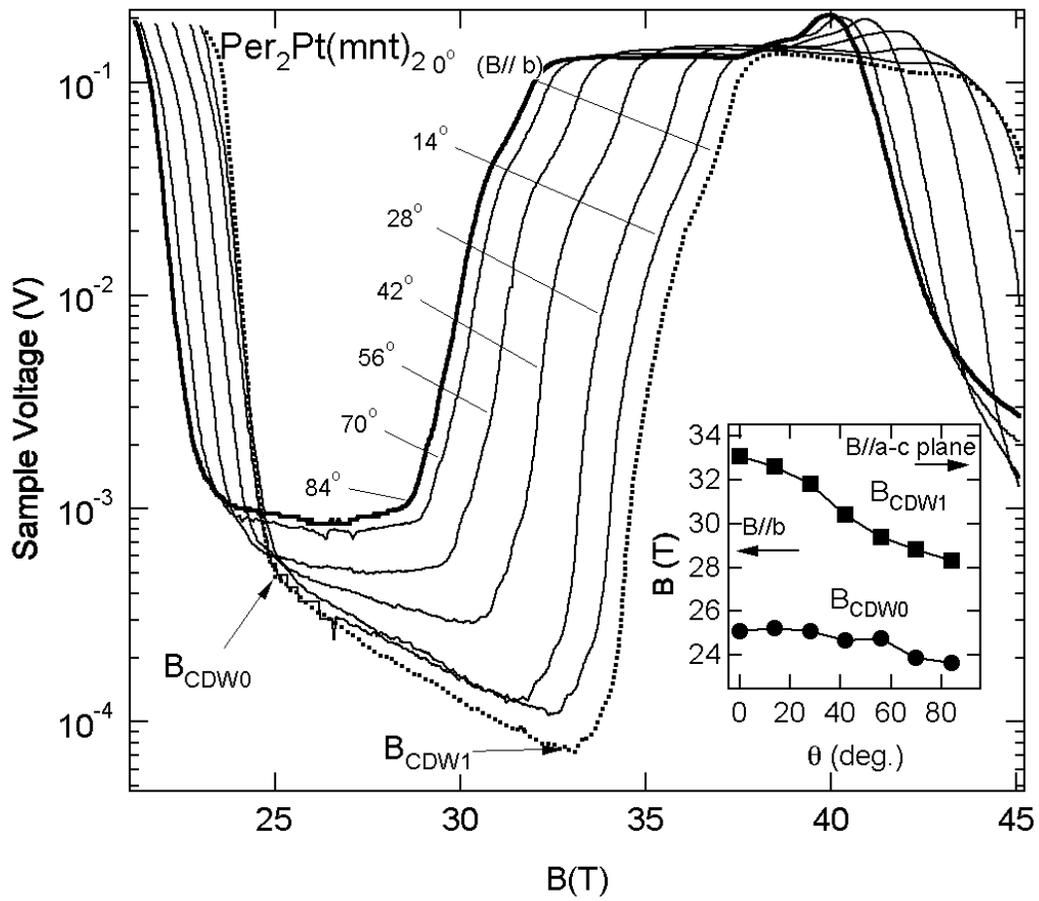

**Figure 4 Graf et al.**

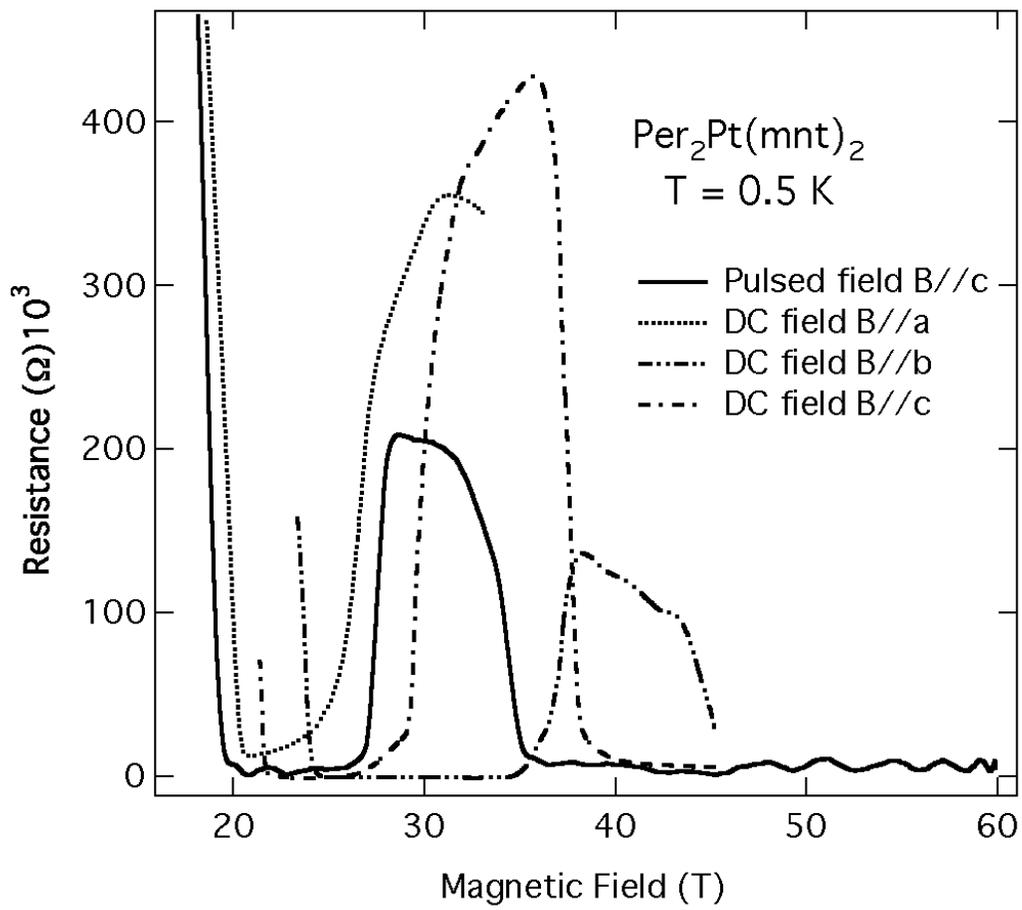

**Figure 5 Graf et al.**